\documentclass[3p, authoryear,12pt]{elsarticle}
\usepackage{graphicx}
\usepackage{amssymb}

\usepackage[utf8]{inputenc}
\usepackage[T1]{fontenc}
\usepackage{graphicx}
\usepackage{amsmath}
\usepackage[breaklinks]{hyperref}
\usepackage{bbm}

\newcommand{\doilink}[1]{\href{http://dx.doi.org/#1}{#1}}

\journal{Journal of Mathematical Psychology, \doilink{10.1016/j.jmp.2013.01.002}}

\begin{document}

\begin{frontmatter}

\title{Information-sharing and aggregation models for interacting minds}

\author[ITP,ICFO]{Piotr Migda{\l}}
\ead{pmigdal@gmail.com}
\ead[url]{http://migdal.wikidot.com/en}
\address[ITP]{Institute of Theoretical Physics, University of Warsaw, Warsaw, Poland}
\address[ICFO]{ICFO--Institut de Ci\`{e}ncies Fot\`{o}niques, 08860 Castelldefels (Barcelona), Spain}
\author[IP]{Joanna Rączaszek-Leonardi}
\address[IP]{Institute of Psychology, Polish Academy of Sciences, Warsaw, Poland}
\author[PUW]{Micha{\l} Denkiewicz}
\address[PUW]{Department of Psychology, University of Warsaw, Warsaw, Poland}
\author[ICM]{Dariusz Plewczynski}
\ead{darman@icm.edu.pl}
\ead[url]{http://cognitivesystems.pl}
\address[ICM]{Interdisciplinary Centre for Mathematical and Computational Modelling, University of Warsaw, Pawi\'{n}skiego 5a, 02-106 Warsaw, Poland}
\date{\today}

\begin{abstract}

We study mathematical models of the collaborative solving of a two-choice discrimination task. We estimate the difference between the shared performance for a group of {\it n} observers over a single person performance. Our paper is a theoretical extension of the recent work of \citet{Bahrami2010} from a dyad (a pair) to a group of {\it n} interacting minds. We analyze several models of communication, decision-making and hierarchical information-aggregation.
The maximal slope of psychometric function
(closely related to the percentage of right answers vs. easiness of the task)
is a convenient parameter characterizing performance. For every model we investigated, the group performance turns out to be a product of two numbers: a scaling factor depending of the group size and an average performance. The scaling factor is a power function of the group size (with the exponent ranging from $0$ to $1$), whereas the average is arithmetic mean, quadratic mean, or maximum of the individual slopes.
Moreover, voting can be almost as efficient as more elaborate communication models, given the participants have similar individual performances. 

\end{abstract}


\end{frontmatter}

\section{Introduction}

Anyone who has ever taken part in group decision making or problem solving has most likely asked themselves at one point or another whether the process actually made any sense.  Would it not be better if the most competent person in the group simply made the decision?  In other words, it is an open question whether a group can ever outperform its most capable member. There have been many studies that have reported group decisions to be less accurate \citep{Corfman1995}. Some studies, however, have concluded that groups --- even when they merely use simple majority voting --- can make better decisions than their individual members \citep{Grofman1978, Kerr2004, Hastie2005}. We ask a more general question: how does the group performance depend upon the individual performances of its participants and the ways in which those participants communicate? 

This question is given new light by recent trends in cognitive psychology, which after a half a century of fascination with isolated cognition in the individual, has finally admitted the individual interaction with the social environment.
It is increasingly understood that joint actions and joint cognition are not limited to situations of committee/voter decisions, but instead, they pervade everyday life and require the constant coordination and integration of cognitive and physical abilities. This new approach, typically called {\it distributed cognition} \citep{Hutchins1995cognition}, or {\it extended cognition} within the social domain \citep{Clark2006}, brings the focus of research to the mechanisms of cognitive and physical coordination \citep{Kirsh2006} that affect this integration. It also brings attention to the comparison of the performance of the group to the performance of the individual. For some tasks that require different types of knowledge and abilities from group participants, groups are likely to outperform individuals \citep{Hill1982}. For other tasks, such as simple discrimination tasks or estimations, a question arises if a group is indeed better than the best of its members. If there are such situations, it is important to know when they arise. 

Group decision making obviously involves members interacting with each other. Casting a vote requires a minimum amount of communication for the individual (only to inform other group members about his or her choice). However, other group decisions allow for extensive communication and negotiations of the decision. Our questions are: 1) which forms of communication are most likely to facilitate an improved outcome, and 2) what is actually being communicated in successful groups? Recent experiments by \citet*{Bahrami2010} have shown that cooperation can be beneficial, even in simple task, and that this benefit is best explained by the participants communicating their relative confidences. In their study, dyads (pairs) performed a perceptual two-choice discrimination task. On every trial participants had to decide which of two consecutive stimuli (sets of Gabor patches) contained a patch with higher contrast. First, decisions were collected from both persons; then, if the decisions were different, the participants were allowed to communicate to reach a joint decision.

The decision data obtained from each person was used to fit a psychometric function, i.e., the probability of that person giving a specific answer, as a function of the difference of the contrast between Gabor patches. These functions describe the person's skill in the task. Similarly, a function describing the skill of the group as a whole can be estimated from the group decisions. As was described by Bahrami: {\it "In experiments (...) psychometric functions were constructed for each observer and for the dyad by plotting the proportion of trials in which the oddball was seen in the second interval against the contrast difference at the oddball location"} (\citet{Bahrami2010}, Supplementary Materials, p. 3).

Various assumptions about the nature of within-group interactions during the joint decision-making process can be made. From these assumptions, we can derive theoretical relationships between the parameters of members' functions and the parameters of the group function. These are the models of decision making. The correctness of each joint decision model can then be tested against empirical data.

\citet{Bahrami2010} described and evaluated four such models. One was own, in which group members communicate their confidence in their individual choices. Another model stemmed from signal detection theory \citep{Sorkin2001}. If members know each other's relative discriminatory ability (i.e., their psychometric functions), the group can make a statistically optimal choice. Thus, under certain conditions, we have an upper bound on group performance. The third model suggested that the dyad is only as good as its best member. Finally, the last model tested was a control model involving random response selection. The study concluded that, when similarly skilled persons meet, they can both benefit from cooperation. A model in which participants communicate their relative confidences best explains this benefit.

We extend the models from \citet{Bahrami2010} to groups of $n$ participants and compare their predictions. Furthermore, we add a model in which a participant either knows the correct answer, or guesses. Importantly, in the case of larger groups, it may be the case that only small subgroups of participants can communicate simultaneously. Thus, we address this issue by considering hierarchical schemes of decision aggregation, in which decisions are first made by subgroups, and then some of these subgroups interact to reach a shared decision.

The paper is organized as follows. In Section \ref{s:discrimination}, we present the \citet{Bahrami2010}  approach to integrating individual discrimination functions in pairs of participants. We use it to assess performance in groups.
In Section \ref{s:communication},  we proceed to formulating a series of models of communication, which express the performance of a group of $n$ persons as a function of their individual performances. In Section \ref{s:aggregation}, we investigate how each model works, assuming several schemes of decision aggregation. Section \ref{s:comparison} compares the introduced models and provides insight into further experimental and theoretical work. Section \ref{s:conclusion} concludes the paper.

\section{Model of discrimination}\label{s:discrimination}

Consider an experiment in which a participant has to make simple discriminatory decisions of varying difficulty. Each trial is assigned a parameter, $c$, that describes the physical distance between stimuli (e.g., in the Bahrami et al. experiment $c$ was the difference in contrast between Gabor patches). Negative $c$ describes a situation in which the right choice is the first of the pair, whereas positive $c$ describes the opposite situation. The absolute value of $c$ reflects the difficulty of a given trial. The lower the value, the more difficult is the resulting trial. 
From now on, we refer to the parameter describing physical difference as {\it stimulus c}. In the case of Bahrami et al. experimental setup, it
can be interpreted as the two-interval stimulus with the difference of contrasts equal to $c$.

By knowing the choices of a certain decision-making agent (in our case either a single participant or a group making the decision together) for a range of stimuli, we can construct a mathematical description of the agent's performance on the task.

For each agent, we can then determine his or her psychometric function: the probability of the agent choosing the second answer as a function of the stimulus, $P(c)$. An ideal responder would be described by the Heaviside step function: $P(c)=0$ for all negative stimuli, and $P(c)=1$ for all positive stimuli (i.e., choosing the second interval if and only if $c>0$). 

Because responders make errors, the actual decision rule and probability are different. 
One way to describe such a response is derived from signal detection theory \citep{Sorkin2001}. According to it, for a stimulus $c$, a participant perceives stimulus $x$, which is a normally distributed random variable centered around $c+b$ and with variance $\sigma$, and decides basing on the sign of $x$. 
Two models described in this paper (Weighted Confidence Sharing and Direct Signal Sharing) use this mechanism explicitly.
The modified realistic decision rule of an agent states that if the observed stimuli $x$ is negative, an
agent decides to select the first patch (therefore interpreting the difference in contrast as negative),
in the case of positive value, the second option is selected.

\begin{figure}[!htbp]
    \centering
        \includegraphics[width=0.60\textwidth]{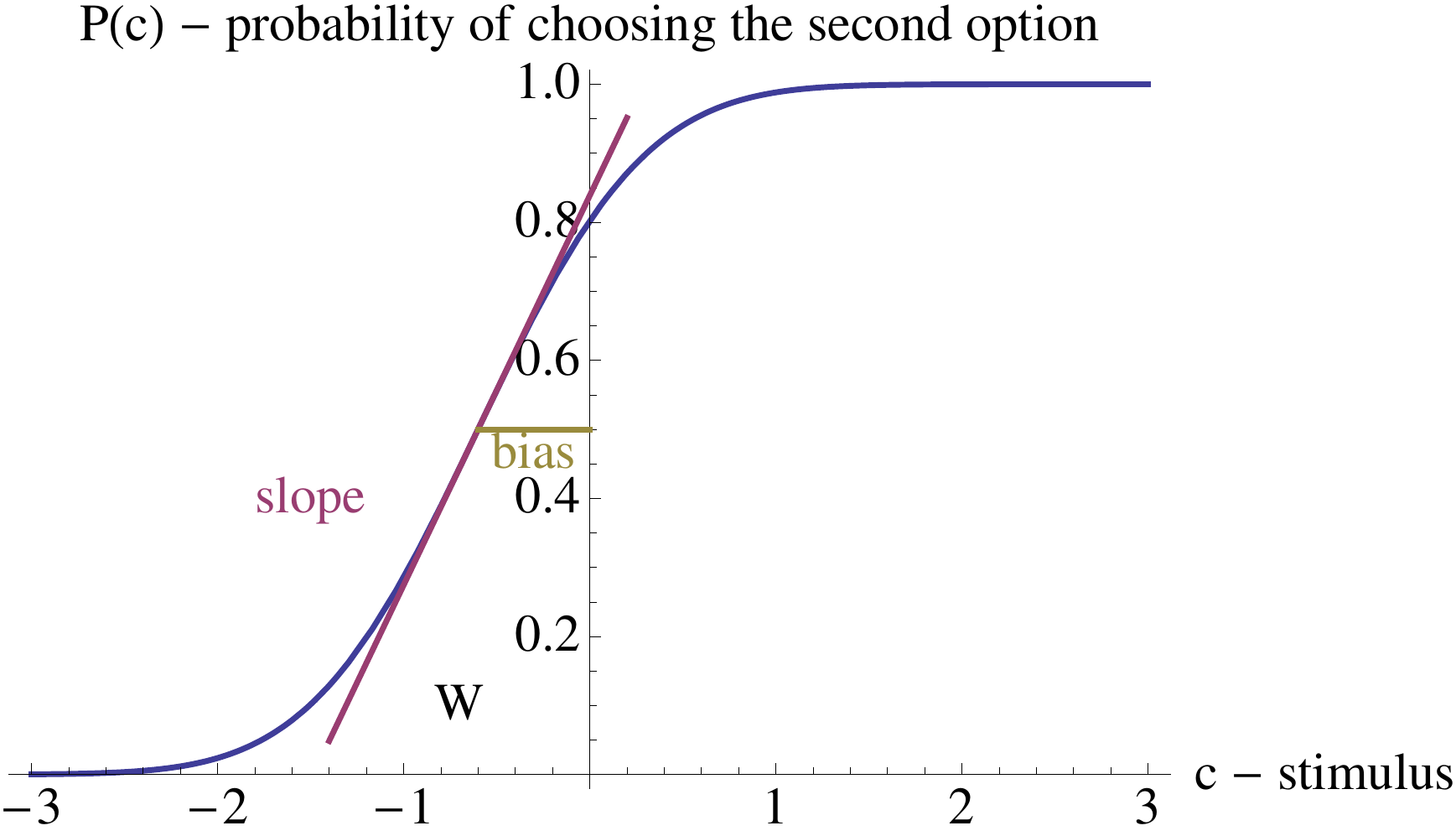}
    \caption{Plot of the psychometric function, with shown slope $s$ and positive bias $b$.
    }
    \label{fig:psychometric}
\end{figure}

In particular, psychometric curves which are  cumulative of the normal distribution: 
\begin{align}
P(c)&=H\left(\tfrac{c+b}{\sigma}\right),\qquad\hbox{ where}\label{eq:pdef}\\
H(x) &= \frac{1}{\sqrt{2\pi}} \int_{-\infty}^{x} \exp\left(-t^2/2\right)dt,
\end{align}
result in a good fit for the experimental data \citep{Bahrami2010}. The parameter $\sigma$ can be interpreted as the participant's uncertainty about the decision. The parameter $b$ is the bias (offset); it represents a tendency to choose a particular answer, see Fig. 
\ref{fig:psychometric}.
The $P(c)$ function, defined as above, can be viewed as a convolution of the step function (the correct answer) and the Gaussian distribution (the discriminative error).

For our purposes, we assume that bias is much smaller than the characteristic width parameter, i.e., $|b|\ll \sigma$. 

Consequently, $\sigma$ becomes the main determinant of the effectiveness of discrimination. It is convenient to choose the maximal slope of the psychometric function 
\begin{align}
s &= \frac{1}{\sqrt{2\pi}\sigma},\label{eq:slopesigma}
\end{align}
as the primary measure of the responding agent's effectiveness.

Now, we can proceed to extending the \citet{Bahrami2010} models. We would like to know how the performance of a group of $n$ people depends upon their individual cognitive performances. Therefore, we need to solve the explicit formulas for the propagation of slopes and biases when combining several responders within each of the different models of communication: 
\begin{align}
s_{model} &= s_{model}(s_1,b_1,\ldots, s_n,b_n),\label{eq:smodel}\\
b_{model} &= b_{model}(s_1,b_1,\ldots, s_n,b_n).\label{eq:bmodel}
\end{align}
Each model is described by the shared decision function 
\begin{align}
P_{model} &= f\left[P_1,\ldots,P_n \right]\label{eq:pmodel},
\end{align}
where $f$ is a functional. For all but two models that we investigate, $P_{model}(c) = f\left[P_1(c),\ldots,P_n(c) \right]$, that is,  the dependence is pointwise (i.e., result for a given $c$ requires only knowing individual $P_i(c)$ for the same $c$). 

We can obtain the effective slope \eqref{eq:smodel} and bias \eqref{eq:bmodel} using straightforward formulas that involve taking the derivative of the psychometric function with respect to the stimulus:
\begin{align}
s_{model} &=\left.P_{model}'(c)\right|_{c=-b_{model}} \approx \left.P_{model}'(c)\right|_{c=0} \label{eq:slopediff}\\
b_{model} & = \left[ b \mbox{ for which } P_{model}(-b)=\frac{1}{2} \right] \approx \frac{P_{model}(0)-\frac{1}{2}}{\left.P_{model}'(c)\right|_{c=0}},\label{eq:biasdiff}
\end{align}
where assuming \eqref{eq:pdef} the approximation for the relative error for both $s$ and $b$ is of order $O(s^2 b^2)$ (or equivalently, $O(\frac{b^2}{\sigma^2})$), where $O(\cdot)$ stands for big $O$ notation. The derivation is in \ref{s:approx}. Note that if $P_{model}(c)$ is a cumulative Gaussian function (as in \eqref{eq:pdef}), then the formulas for slope \eqref{eq:slopesigma} and \eqref{eq:slopediff} are equivalent. However, it can be used as a {\it definition} of the slope and the bias in the general case of an arbitrary communication strategy $P_{model}(c)$, even if \eqref{eq:pdef} does not hold. Bear in mind that for practical applications, we expect $P_{model}(c)$ to be close enough to the cumulative Gaussian function. Moreover, when there are no biases, for all decision-making considered in this paper, the maximal slope is at $c=0$. 

A question arises about the relation between the psychometric curve parameters and the expected rate of errors. To assess the average amount of incorrect answers we could expect from a responder, we introduced the following quantity, 
\begin{align}
W(\sigma,b) &= \int_{-\infty}^{0}P(c)dc + \int_{0}^{\infty} \left[ 1-P(c)\right]dc\\ 
 &= \sqrt{\frac{2}{\pi}}\sigma \exp\left(-\tfrac{b^2}{2\sigma^2}\right) + b\left[ 2H\left(\tfrac{b}{\sigma}\right)-1 \right]\label{eq:penaltyres},
\end{align}
where we integrated the error function \citep{AbramowitzStegun1965}.
For a  uniform distribution and range of stimuli, $(-r,r)$ for $r \gg (\sigma + |b|)$, the rate of the incorrect responses is given by $W(\sigma,b)/(2r)$. The average number of wrong answers is always reduced when lowering either width or bias, regardless of the other parameter's value. This fact further justifies the choice of the slope as the proper effectiveness measure. When there is no bias, \eqref{eq:penaltyres} simplifies to $W(\sigma,0)=2/s$; thus, the rate of the incorrect responses is $1/(r s)$.

\section{Information-sharing models}\label{s:communication}

In this section, we discuss different models of information sharing for $n$ participants. It is important to underline that the models incorporate the process of perceiving (what the subjects may know), the state of mind (what the subjects know), and the communication and the decision-making process (usually Bayes-optimal). We briefly define the assumptions of each model and justify it in psychological terms. We give results in terms of the effective psychometric function, $P_{model}(c)$, the effective slope, $s_{model}$, and sometimes the 
effective bias, $b_{model}$ (as for a few models the bias is poorly defined). Whenever calculations of $P_{model}(c)$ are not straightforward, we give some insight into the underlying mathematics.

We investigate the following models:
\begin{itemize}
\item \ref{s:rr} Random Responder,
\item \ref{s:vot} Voting,
\item \ref{s:bd} Best Decides,
\item \ref{s:wcs} Weighted Confidence Sharing,
\item \ref{s:dss} Direct Signal Sharing,
\item \ref{s:tw} Truth Wins.
\end{itemize}

\subsection{Random Responder}\label{s:rr}

\paragraph*{Model}

The trial decision of a random group member is taken as the group decision.

\paragraph*{Motivation}
Random Responder serves as one of the reference models, and it is not expected to be fulfilled in most of realistic settings. Random factors determine the collective decision, i.e., communication is seen as ineffective within framework of this model. Sometimes the decision is not based on any evidence and people may have very misleading impressions of their own accuracy. Additionally, their decisions may depend more upon a group member's charisma or persuasive skills than his or her psychometric skills. In the work of \citet{Bahrami2010}, this model is called 'Coin flip'. 

\paragraph*{Results}

\begin{align}
P_{RR}(c) &= \frac{1}{n}\sum_{i=1}^{n} P_i(c)
\end{align}
After the differentiation, one obtains the slope \eqref{eq:slopediff} and the bias \eqref{eq:biasdiff}:
\begin{align}
s_{RR} & \approx \frac{s_1+\ldots+s_n}{n}\\
b_{RR} & \approx \frac{s_1b_1+\ldots+s_n b_n}{s_1+\ldots +s_n} 
\end{align}
The relative error both for $s_{RR}$ and $b_{RR}$ is $O(s_1^2 b_1^2) + \ldots + O(s_n^2 b_n^2)$.
Note that $P_{RR}(c)$ is not normal \eqref{eq:pdef}.

\subsection{Voting}\label{s:vot}

\paragraph*{Model}
Each participant makes her or his own decision. The majority vote determines the decision of the group. In a case of equal votes for two outcomes, a coin is flipped. 
\paragraph*{Motivation}
People may have no access to their accuracy (or they cannot communicate it reliably); thus, a good strategy is to take voting as the final consensus result.
\paragraph*{Results}

\begin{align}
&P_{Vot}(c) = \sum_{k=1}^{\lfloor \frac{n-1}{2} \rfloor} \sum_{\vec{i}} \left[ 1-P_{i_{1}}(c)\right]\cdots \left[ 1-P_{i_{k}}(c)\right] P_{i_{k+1}}(c)\cdots P_{i_{n}}(c)\\
 & +\left[ \frac{1}{2} \sum_{\vec{i}} \left[ 1-P_{i_{1}}(c)\right]\cdots  \left[ 1-P_{i_{n/2}}(c)\right] P_{i_{n/2+1}}(c)\cdots P_{i_{n}}(c)  \right]_{\mbox{if $n$ is even}},\nonumber
\end{align}
where sum over $\vec{i}$ denotes sum over every permutation of participants. We obtain (derivation in \ref{s:appvot})
\begin{align}
s_{Vot} &\approx
\frac{s_1+\ldots+s_n}{n}\times
\left\{
\begin{matrix}
\frac{n}{2^n} {n \choose n/2} & \hbox{if $n$ is even} \\
\frac{n}{2^{n-1}} {(n-1) \choose (n-1)/2} & \hbox{if $n$ is odd} 
\end{matrix}
\right.\\
 &\approx \sqrt{\tfrac{2}{\pi}} \times \sqrt{n} \times \frac{s_1+\ldots+s_n}{n}\label{eq:svotas}\\
b_{Vot} &\approx \frac{s_1b_1+\ldots+s_n b_n}{s_1+\ldots +s_n}
\end{align}
The  $P_{Vot}(c)$ is not normal \eqref{eq:pdef}. The relative error both for $s_{Vot}$ and $b_{Vot}$ is $O(s_1 b_1) + \ldots + O(s_n b_n)$. 
Note that the addition of an odd member to a group does not increase its average performance. The formula \eqref{eq:svotas} is an asymptotic expression for large $n$, which utilizes the Wallis formula. For $n=2$, the Random Responder and Voting models yield the same results.

\subsection{Best Decides}\label{s:bd}

\paragraph*{Model}
The most accurate member of the group makes the decision. This model is called {\it Behavior and Feedback} in \citet{Bahrami2010}. In this model, we will focus on the case with no bias, $b=0$. Nonzero bias would make the result difficult to state in explicit form; see \eqref{eq:penaltyres} for further explanation.

\paragraph*{Motivation}
In some experimental settings, members of the group can determine, who is the most accurate (e.g., when feedback is present). Group members can then let that individual make the final decision. Studies by \citet{Henry1995} suggest that, at least in some types of tasks, participants can identify the most proficient member, so our assumption is plausible. As in the previous models, there is no (effective) communication between the members of the group.

\paragraph*{Results}

\begin{align}
P_{BD}(c) &= P_{\mbox{member with the highest $s$}}(c)\\
s_{BD} &= \max(s_1,\ldots,s_n)
\end{align}
When biases are large, the group psychometric function is that of the most effective participant (i.e., one with the lowest $W(\sigma_i,b_i)$ \eqref{eq:penaltyres}), $P_{BD}(c)=P_i(c)$. This strategy is most beneficial for a group with very diverse individual performances.

\subsection{Weighted Confidence Sharing}\label{s:wcs}

\paragraph*{Model}
Group members share their relative confidences $z_i=x_i/\sigma_i$. The group decision depends on the sign of $\sum_{i=1}^n z_i$, i.e., for the negative they choose the first option and for the positive they choose the second. This model requires each $P_i(c)$ to be normal \eqref{eq:pdef}.
\paragraph*{Motivation}

The value $x_i$ is the stimulus perceived by $i$-th participant and has a distribution with density $P_i'(c)$, as it is in \citet{Sorkin2001}. We assume the confidence to be a continuous variable. The true stimulus $c$ is, of course, common for all participants in a given trial. The relative confidence is equivalent to a $z$-score, if the participant is unbiased (i.e., it is related to the probability that the participant is right). Put differently, participants know their $z$-scores on a given trial but are unaware of their own parameters $s$ and $b$. This model was first introduced by \citet{Bahrami2010}. It is possible that, in an experimental trial, each participant can estimate and effectively communicate their relative confidences, by using a coarse real-world approximation of one's $z$-score, e.g., 'I lean towards 1st', or 'I am almost sure it is the 2nd' \cite{Fusaroli2012}. The study by \citet{Bahrami2010} suggests that this model most accurately describes dyad performance.

Given relative confidences $\vec{z}=(z_1,\ldots,z_n)$, the group has to determine whether to choose the first or the second option. If there are only two participants with different opinions, the one with the stronger confidence (for a given trial) decides. This can be written as follows: the group chooses the first option if $z_1+z_2 \leq 0$, the second option otherwise, yielding an optimal strategy \citep{Bahrami2010}. In the general case of $n$ participants, we use the Bayes optimal reasoning. We calculate the probability that the stimulus is positive (and thus the second answer is correct) given the $z$-scores provided by each participant: 
\begin{align}
p(c>0|\vec{z}) = \int_{c=0}^\infty p(c|\vec{z})dc
= \frac{\int_0^\infty p(\vec{z}|c)p(c)dc}{\int_{-\infty}^\infty p(\vec{z}|c)p(c)dc},\label{eq:whenposit}
\end{align}
where $p(c)$ is the probability of a discrimination task with $c$. The probability of observing $z_i$-score, given stimulus $c$, is $P_i'(c- \sigma_i z_i)$. Thus
\begin{align}
p(\vec{z}|c) = P_1'(c-x_1)\cdot\ldots\cdot P_n'(c-x_n).\label{eq:pzc}
\end{align}
Let us assume that the displayed stimulus has a uniform distribution, i.e., that $p(c)$ is constant (not going into mathematical nuances). To define the decision function, we need to know when $p(c>0|\vec{z}) \geq 1/2$ or, in other words, when the probability that the second answer is correct is greater than $1/2$. As \eqref{eq:pzc} is a Gaussian function of $c$, finding its maximum leads to the condition
\begin{align}
\frac{x_1}{\sigma_1^2}+\ldots+\frac{x_n}{\sigma_n^2}&\geq 0,
\end{align}
or equivalently, using the slope parameter,
\begin{align}
s_1 z_1+\ldots+s_n z_n&\geq 0.
\end{align}
Thus, when the condition holds, choosing the second option is the Bayes optimal choice. Unfortunately, in this model we only have access to values of $\vec{z}$, not to individual performances. To obtain the precise answer, we need to know the whole distribution of $\sigma_i$ (or $s_i$). Instead, we can use the approximate condition for the choice of the second option,
\begin{align}
z_1+\ldots+z_n&\geq 0\label{eq:zsum},
\end{align}
to obtain a lower bound on the performance. The condition is exact for participants with equal performances (and should be close to the optimal if the values of $\sigma_{i}$ do not vary much). This equation can be seen as a type of a weighted voting, where weights depend on subjective confidences, but not on individual performances. Members do not know their own --- or their peers' --- performance scores, so there is no justification for assigning more or less weight to a particular member throughout the experiment. The only thing that matters is each member's confidence in the present trial.

\paragraph*{Results}

To calculate $P_{WCS}(c)$, we need to compute, given stimulus $c$, the probability of obtaining set $\vec{z}$ with a positive sum \eqref{eq:zsum}. Thus 
\begin{align}
P_{WCS}( c) &= \int_{x_1/\sigma_1 + \ldots + x_n/\sigma_n \geq 0}  \exp \left[ - \frac{(c+b_1-x_1)^2}{2\sigma_1^2}+\right.\\
&\left.-\ldots- \frac{(c+b_n-x_n)^2}{2\sigma_n^2}\right]\frac{dx_1\cdots dx_n}{(2 \pi)^{n/2} \sigma_1 \cdots \sigma_n}\nonumber\\
&= H\left[ \sqrt{2\pi} s_{WCS} \left( c + b_{WCS}\right)  \right],
\end{align}
where the integration is based upon the fact that a sum of Gaussian random variables $z_i$ is a Gaussian random variable \citep{Piau2011}. The resulting parameters are:
\begin{align}
s_{WCS} & = \sqrt{n} \times \frac{s_1+\ldots+s_n}{n}, \\
b_{WCS} & =\frac{s_1 b_1+\ldots+s_n b_n}{s_1+\ldots+s_n}.
\end{align}

Again, note that the above result for $s_{WCS}$ is the lower boundary value for optimal Bayesian reasoning, exact only for $n=2$ (due to symmetry) and a group of participants with the same performance. By knowing the exact distribution of individual performances, we can obtain a better (or at least the same) group performance. Then, instead of the summation of individual $z$-scores \eqref{eq:zsum}, one will get a more complicated formula for the decision.

\subsection{Direct Signal Sharing}\label{s:dss}

\paragraph*{Model}
Group members share both their perceived stimuli $x_i$ and their $\sigma_i$. The group decision depends on the sign of $\sum_{i=1}^n x_i/\sigma_i^2$.  This model requires each $P_i(c)$ to be normal \eqref{eq:pdef}.
\paragraph*{Motivation}
As for the WCS, we assume that the value $x_i$ is the stimulus perceived by $i$-th participant and has a distribution with the density $P_i'(c)$, as it is in \citep{Sorkin2001}. The group possesses complete knowledge about the characteristics of its members and their perceptions, so its effectiveness is hindered only by the skill of the participants, not by communication. This model constitutes the upper bound for group performance, provided that the stimuli are fully defined by their stimulus values (and perceived according to the discussed model). In the case of a more complex, non-perceptive task, it is possible for a group to exceed this bound \citep{Hill1982}. For example, this could occur when participants' skills complement each other. People know the strength of the stimuli but also their own sensitivity. If the feedback is provided, we can plot $x$ versus $c$ to get $\sigma$.

\paragraph*{Results}
The final group decision follows the standard derivation of n classifiers collecting independent results with normal distribution (e.g., \citet{Sorkin2001} and \citet{Bahrami2010}):
\begin{align}
P_{DSS}(c) &= \frac{1}{\hbox{normalization}}\int_{-\infty}^{c} P_1'(x)\cdot\ldots\cdot P_n'(x) dx\\
s_{DSS} &= \sqrt{s_1^2+\ldots+s_n^2} = \sqrt{n} \times \sqrt{\frac{s_1^2+\ldots+s_n^2}{n}}\\
b_{DSS} &= \frac{s_1^2 b_1+\ldots+s_n^2 b_n}{s_1^2+\ldots+s_n^2}
\end{align}
Note that, regardless of the distribution of the individual performances, the group performance outscores both Best Decides and Weighted Confidence Sharing.

\subsection{Truth Wins}\label{s:tw}

\paragraph*{Model}
We assume that on each trial each member is in one of the two states: either they know the right answer or they are aware of their own ignorance. In the latter case, a random guess is made. It is sufficient to have a single group member perceive the stimuli correctly to get the correct group answer. We assume no bias, as there is no possible way to treat it consistently and it introduces false convictions. 

\paragraph*{Motivation}
For so-called Eureka problems, the signal-theoretic limit can be exceeded \citep{Hill1982}. The key is that the answer to such a problem has the property of demonstrability: it allows a single member who has the correct answer to easily convince the rest of the group of its correctness \citep{Laughlin1975}.
People know if they see the 'right' stimuli (and all errors are due to guessing, not to false observations). This model has received much attention in group decision theory, e.g., in \citet{Davis1973}. It is appropriate in situations when the correctness of a solution can be demonstrated. However, we do not expect this model to be applicable to tasks similar to that of \citet{Bahrami2010}. This model serves as a control and an explicit example of a result beyond one provided by the Direct Signal Sharing model. We included it with the aim of generalizing the models to different decision situations.

\paragraph*{Results}
The probability that the responder knows with certainty the right answer is
\begin{align}
R(c) = \left| 2 P(c) - 1 \right|.
\end{align}
That is, we have a reversed formula saying that, when a responder knows the answer with probability $R(c)$, the responder answers correctly with probability $R(c)+(1-R(c))/2$ (as there is the chance to answer correctly by a random guess). The probability that at least one person knows the correct answer is 
\begin{align}
R_{TW}(c) &= 1 - \left[ 1-R_1(c)\right]\cdot\ldots\cdot\left[ 1-R_n(c)\right].
\end{align}
Consequently, 
\begin{align}
P_{TW}(c) &= \frac{\mbox{sign}(c)R_{TW}(c) +1}{2}\\
s_{TW} & = n \times \frac{s_1 + \ldots + s_n}{n},
\end{align}
where the slope is a result of straightforward differentiation \eqref{eq:slopediff}.

This model yields much better results than other models; note, however, that the absence of false observations is a strong requirement. Other models have to operate without this assumption. Note that the $P_{TW}(c)$ is not normal.

\section{Aggregation of information in hierarchical schemes}\label{s:aggregation}

So far, we have assumed that information from all participants is simultaneously collected and used in the group decision. One may argue that this is unrealistic for human communication in groups of more than a few persons. We, therefore, propose hierarchical models (schemes) in which only small subgroups can communicate at a particular time. Each of these subgroups reaches its own decision, in a manner described by one of the models introduced in the previous section. Hence, the subgroup can be regarded as a decision-making agent, described by a slope and a bias. The subgroup can then communicate with other subgroups or individual members, which results in larger groups being created, until all information is gathered and the final decision is made.

The results of employing a multi-level decision system can significantly deviate from what simultaneous information collection predicts. For instance, in a two-level voting system, which has been widely studied in the context of election results \citep{Davis1973, Laughlin1975} the final outcome depends heavily upon the distribution of votes in the subgroups, sometimes allowing minority groups to overcome the majority, sometimes exaggerating the power of the majority. It is thus interesting to study the possible effects of such hierarchical systems.

We propose the following model for communication of $n$ participants: 
\begin{enumerate}
\item In the beginning there are $n$ agents.
\item Each turn only $g$ (for our purpose: $2$ or $3$) agents (groups or individuals) share their information according to a chosen model. These agents are then merged into one agent (defined by $s_{model}(s_1,\ldots,s_g)$).
\end{enumerate}
In other words, a group of people who shared information, is treated as a single agent in the next turn. There are two free parameters: 
\begin{itemize}
\item The model used to combine members' parameters into group parameters.
\item How the groups are formed, i.e., the way to determine which agents should interact in given turn. 
\end{itemize}

\begin{figure}[!htbp]
    \centering
        \begin{tabular}{cc}
        \includegraphics[width=0.45\textwidth]{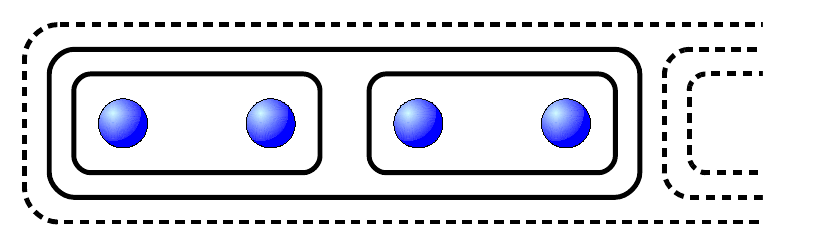} &
        \includegraphics[width=0.45\textwidth]{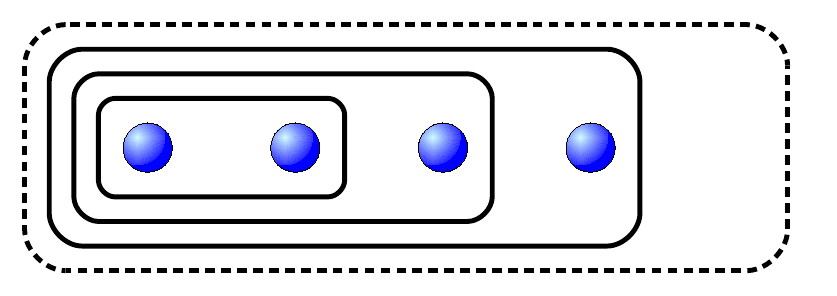}\\ 
        Shallow Scheme & Deep Scheme
        \end{tabular}
    \caption{Diagram of the interaction ordering for aggregation schemes for $g=2$: Shallow Scheme --- each turn two agents from the least numerous groups interact, Deep Scheme --- each turn a single participant joins the previously formed group.}
    \label{fig:aggregab}
\end{figure}

Let us consider the following ways in which groups can form (see Fig. \ref{fig:aggregab} for the 
diagram of the two first schemes): 
\begin{itemize}
\item \ref{s:ha} Shallow hierarchy: Each turn $g$ agents from the groups with the least number of participants interact.
\item \ref{s:hb} Deep hierarchy: Each turn $g-1$ agents join to the group with largest number of participants (that is, there is only one group to which each turn $g-1$ agents join).
\item \ref{s:hc} Random hierarchy: Each turn $g$ random agents interact.
\end{itemize}
Above, by participants we understand the total number of individuals that were merged into an agent.

For some models, the way in which groups are formed is irrelevant for obvious reasons. This is the case for Random Responder, Best Decides, Direct Signal Sharing and Truth Wins. The result is always the same and is equivalent to the simplest situation without any hierarchy. The models, which are affected to some degree, are as follows: Weighted Confidence Sharing and Voting.

Note that, in principle, agents do not know their own slopes, so the order of interactions cannot depend upon the individual (or group) $s_i$.  However, as  both $s_{WCS}$ and $s_{Vot}$ depend linearly on $s_i$, averaging over every permutation of participants yields a result that is proportional to the arithmetical mean of $s_i$, or $\langle s \rangle$. Consequently, to investigate the influence of the hierarchical information-aggregation, it is sufficient to treat each participant as if his/her performance is equal to $\langle s \rangle$.

For convenience, we consider a more general model with the parameter  (the amplification multiplier) depending on $g$ (the group size) as follows: 
\begin{align}
s_{a_g}(s_1,\ldots ,s_g) = a_g \frac{s_1+\ldots+ s_g}{g}.\label{eq:sag}
\end{align}
This generalization describes both WCS ($a_2=\sqrt{2}$, $a_3=\sqrt{3}$, $\ldots$) and Voting ($a_3 = 3/2$, $\ldots$), and it allows us to give results in an elegant general form.

\subsection{Shallow hierarchy}\label{s:ha}

Our justification for the Shallow hierarchy is the following: people may locally find their partners and then make a collective decision. Then, iteratively, groups of the same (or similar) size make the collective decision.

The analysis is simple when the number of participants is a power of $g$, i.e., $n=g^k$, where $k$ is a natural number.Then, every several elementary steps the number of agents is reduced by the factor of $g$, and agents' slopes are multiplied by the factor $a_g$. In the end, we get
\begin{align}
s_{a_g,Shallow,g} = \left(a_g\right)^k \langle s \rangle = n^{\log_g(a_g)} \langle s \rangle.
\end{align}

In particular, for Weighted Confidence Sharing (i.e. $a_g=\sqrt{g}$), we reach the saturation
\begin{align}
s_{WCS,Shallow,g} = \sqrt{n} \langle s \rangle. \label{eq:wcsag}
\end{align}
Thus, the aggregation process does not introduce a decrease in the group performance when it is compared to collecting all information at once. The formula \eqref{eq:wcsag} holds only for $n$ that is a power of $k$. However, for different $n$s the formula works as a very good approximation. See Fig. \ref{fig:aggreg} for the numerical results. The  relation (i.e., that  for groups of size $n=g^k$ we reach the efficiency of model without aggregation or $s_{a_g,Shallow,g}=s_{a_g,Shallow}$) is true for every model described by \eqref{eq:sag} with $a_g = g^\alpha$ for any $\alpha$.  

In the Voting model we need to consider the aggregation in a group of at least three (i.e., $g=3$ and $a_g=3/2$). Otherwise, it is equivalent to the Random Responder model. For $n$ being the power of three we get, 
\begin{align}
s_{Vot,Shallow,g=3} = n^{\log_{3}3/2} \langle s \rangle \approx n^{0.37} \langle s \rangle,
\end{align}
which works as a good approximation also for the general odd $n$. For every even $n$, there is at least one process with two parties, which significantly decreases the total performance (as voting for two participants reduces to a coin flip).

\subsection{Adding one or two at a time}\label{s:hb}

In this case, there is a single group to which single agents join one after another. The resulting slope is as for the Weighted Confidence Sharing model: 
\begin{align}
s_{WCS,Deep,g=2}&=2^{-(n-1)/2} \langle s \rangle+\sum_{i=1}^{n-1} 2^{-i/2} \langle s \rangle =\left(1+\sqrt{2}-2^{1-n/2}\right) \langle s \rangle
\end{align}
and for the Voting model for an odd $n$ and aggregation of three
\begin{align}
s_{Vot,Deep,g=3} &= 2^{(n-1)/2} \langle s \rangle +2\sum_{i=1}^{(n-1)/2} 2^{-i} \langle s \rangle =\left(2-2^{-(n-1)/2} \right) \langle s \rangle
\end{align}
We see that the Deep hierarchy is very inefficient. The multiplier of $\langle s \rangle$ converges to a constant. This leads to the conclusion that simultaneous aggregation (i.e., Shallow hierarchy) is not only more natural but also much more efficient.

To obtain the asymptotic value of $s_{a_g,Deep,g} $, we can consider an equilibrium situation wherein $g-1$ individuals join the group, which has already reached the limit 
\begin{align}
s_{a_g,Deep,g} &= a_g \left(\frac{g-1}{g} \langle s \rangle + \frac{1}{g}s_{a_g,Deep,g} \right).
\end{align}
This leads to: 
\begin{align}
s_{a_g,Deep,g} &= \frac{g-1}{g/a_g -1}\langle s \rangle.
\end{align}

\subsection{Random hierarchy}\label{s:hc}

What happens between the Shallow hierarchy and the Deep hierarchy?  If the groups merge at random, is the final $s$ closer to the most efficient aggregation scheme, or to non-scaling (e.g., adding a  few members at a time)?  The answer, not surprisingly, lies in between these two extremes.

We parameterize time with $t$ starting from $0$. Each turn $g$ agents merge into one of the slope \eqref{eq:sag}. The current number of agents is described by $n_t=n_0-(g-1)t$. We investigate how the distribution of slopes $\rho_t(s)$ evolves with time, which reads
\begin{align}
&\rho_{t+1}(s) - \rho_t(s) =\label{eq:rhoevoldis}\\
 &-g\frac{\rho_t(s)}{n_t} + \int \frac{\rho_t(s_1)}{n_t} \cdots \frac{\rho_t(s_g)}{n_t} \delta\left( s_{model}(s_1,\cdots,s_g) - s \right) ds_1\cdots ds_g,\nonumber
\end{align}
where $\delta$ is the Dirac delta, i.e., a distribution such that $\int_{-\infty}^{\infty} f(x) \delta(x-x_0) dx = f(x_0)$. The difference in distributions $\rho_{t+1}(s) - \rho_t(s)$ involves two processes. The first expression means that we take $g$ random agents. These agents interact and are removed from the distribution. The second expression means that, for every possible group of $g$ agents (with slopes $s_1,\ldots,s_n$), a new agent is created with the slope $ s_{model}(s_1,\ldots,s_g)$.

\begin{figure}[!htbp]
    \centering
        \begin{tabular}{cc}
        \includegraphics[width=0.45\textwidth]{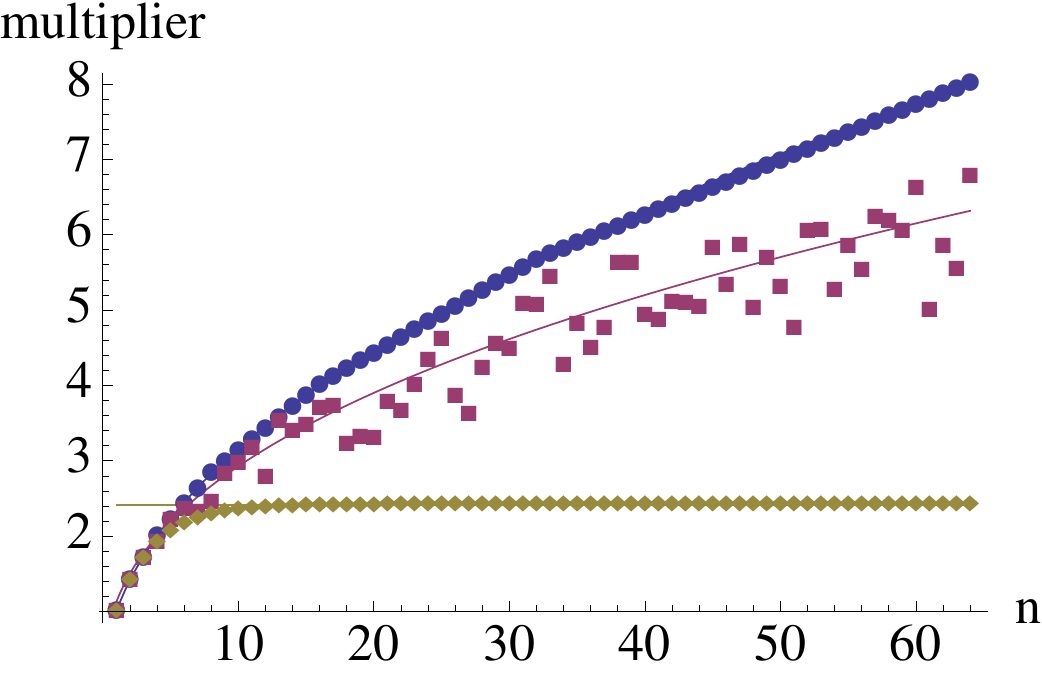} &
        \includegraphics[width=0.45\textwidth]{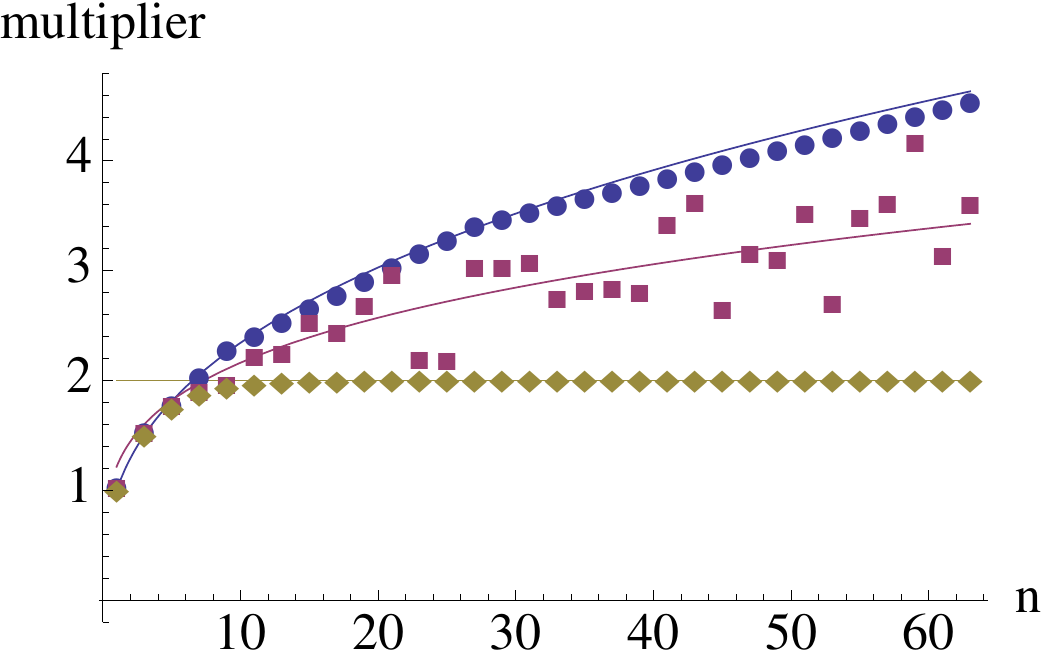}\\ 
        Weighted Confidence Sharing, $g=2$ & Voting, $g=3$
        \end{tabular}
    \caption{Plot of numerically obtained multipliers of $\langle s \rangle$ for models with aggregation of information. Weighted Confidence Sharing with $g=2$ for aggregation hierarchies: Shallow (circles), Deep (diamonds) and Random (squares). Voting with $g=3$, and only for odd number of participants, for aggregation hierarchies: Shallow (circles), Deep (diamonds) and Random (squares). The lines are the respective analytical results from Sec. \ref{s:aggregation}. The numerical results for the Random hierarchy are taken from one shot, i.e., they are not averaged.}
    \label{fig:aggreg}
\end{figure}

Note that we use integrals, but sum over a finite set will give the same result. The parameter we are most concerned with is the mean slope, that is 
\begin{align}
\langle s \rangle_t = n_t^{-1} \int s \rho_t(s) ds.
\end{align}
We multiply \eqref{eq:rhoevoldis} by $s$ and integrate $\int \cdot ds$. In our case, \eqref{eq:sag}, this gives a relatively simple result:
$n_{t+1}\langle s \rangle _{t+1} = n_{t}\langle s \rangle_t - g \langle s \rangle_t + a_g \langle s \rangle_t$ or
\begin{align}
\langle s \rangle _{t} = \frac{n_0 - (g-1)t+a_g-1}{n_0-(g-1)t} \langle s \rangle _{t-1}.
\end{align}
To obtain the final result, we need to calculate $\langle s \rangle _{t_{max}}$ at the point of time when only one agent remains.
We consider $t_{max}=(n_0-1)/(g-1)$ to be an integer (e.g., for $g=3$ we need to consider an odd number of participants, for $g=2$ there are no restrictions). Then, remembering that $\langle s \rangle_0 = \langle s \rangle$ and $n_0 = n$, we get
\begin{align}
s_{a_g,Random,g} &= \prod_{t=1}^{t_{max}} \left( \frac{n_0 - (g-1)t+a_g-1}{n_0-(g-1)t} \right) \langle s \rangle\\
&=\frac{\Gamma\left( \frac{1}{g-1}\right)}{\Gamma\left( \frac{a_g}{g-1} \right)} 
\frac{\Gamma\left( \frac{n_0}{g-1}+\frac{a_g-1}{g-1} \right)}{\Gamma\left( \frac{n_0}{g-1} \right)} \langle s \rangle\\
&\approx \frac{\Gamma\left( \frac{1}{g-1}\right)}{\Gamma\left( \frac{a_g}{g-1} \right) (g-1)^{(a_g-1)/(g-1)}} \times n^{(a_g-1)/(g-1)} \times \langle s \rangle
\end{align}
where $\Gamma(x)$ is the Euler gamma function, and we applied the Stirling approximation. For $g=2$, we obtain the neat result
\begin{align}
s_{a_g,Random,g=2} \approx \frac{1}{\Gamma(a_2)} n^{a_2-1} \langle s \rangle,
\end{align}
in particular, for the Weighted Confidence Sharing model ($a_2=\sqrt{2}$) we get
\begin{align}
s_{WCS,Random,g=2} \approx 1.13 n^{0.41} \langle s \rangle,
\end{align}
whereas for the Voting model for $g=3$ (and odd number of participants) we get
\begin{align}
s_{Vot,Random,g=3} \approx 1.22 n^{0.25} \langle s \rangle.
\end{align}
In Fig. \ref{fig:aggreg} , we present plots for Weighted Confidence Sharing in aggregation groups of two, and Voting in groups of three. We use both analytical approximations and numerical results.

\section{Discussion on results and comparison of models}\label{s:comparison}

\begin{figure}[!htbp]
    \centering
        \includegraphics[width=0.90\textwidth]{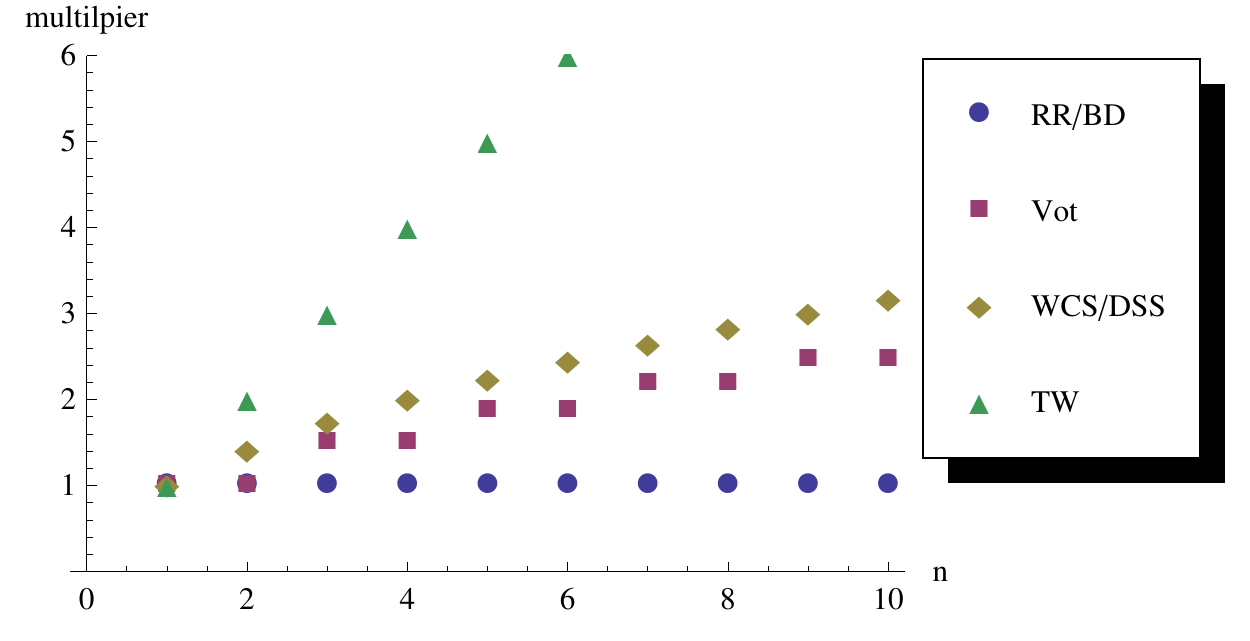}
    \caption{Plot summarizing multipliers for different models.}
    \label{fig:multi}
\end{figure}

For each investigated model, we arrived at the formula for the slope of the group as a function of individual slopes,
\begin{align}
s_{model}(s_1,\ldots,s_n) = \hbox{multiplier}_{model}(n) \times \hbox{mean}_{model}(s_1, \ldots, s_n).
\end{align}
Explicit results can be found in Tab. \ref{tab:cns} and Fig. \ref{fig:multi}. Note that the formula is a product of two quantities --- performance as a function of the group size (i.e., the multiplier), and the mean of the individual slopes (if the better-performing contribute more to the outcome). For equally skilled participants, only the multiplier matters, whereas for a group of people with high variance in performance, the type of mean is crucial.

We not only solved the problem for a particular list of models, but we also constructed a general framework for the collaborative solving of a two-choice task, i.e., the group performance can be written as 
\begin{align}
s_{model}(s_1,\ldots,s_n) = d \times n^\alpha \times \left( \frac{s_1^p + \ldots + s_n^p}{n} \right)^{1/p},\label{eq:s_model_fit}
\end{align}
where parameters $d$, $\alpha$ and $p$ can be fitted for any experimental data, even data not covered by the models we investigated. Note that for $p=1$ we arrive at the arithmetic mean, for $p=2$ we arrive at the quadratic mean, and $p\rightarrow \infty$ we arrive at the maximum. For the models we investigated, \eqref{eq:s_model_fit} is either an exact solution (RR, WCS, BD, DSS, TH) or a good approximation (Voting, information aggregation schemes).  If the result is exact, then $d=1$ (to be consistent with the case of $n=1$).

For a given list of slopes $(s_1,\ldots,s_n)$, it is possible to write relations with the performances (slopes) for different models which read as follows: 
\begin{align}
s_{RR} \leq s_{Vot} < s_{WCS} \leq s_{DSS} \leq s_{TW}.
\end{align}
An average-performing participant is expected to benefit from participating in a joint task, unless the responder is chosen at random (in which case there is neither a gain nor a loss). It is somewhat more difficult to compare the Best Decides model to the other models, as it highly depends on the distribution of the participants' skills. We can write 
\begin{align}
s_{RR} < s_{BD} < s_{DSS} \leq s_{TW}.
\end{align}
However, how does the Best Decides model relate to the Voting and the Weighted Confidence Sharing models?  The answer lies in the comparison of the most skilled participant with the average performance, i.e., $\hbox{max}(s)/\langle s \rangle$. If this ratio is greater than $\approx 0.8 \sqrt{n}$, the Best Decides model outperforms the Voting. If the ratio is greater that $\sqrt{n}$, Best Decides outperforms the WCS as well.  For example, when there is one expert (with $s_{exp}>1$ among $s_{non-exp}=1$) among the total number of $n$ participants, then only when $s_{exp}>\sqrt{n}+1$ it is better for a group to use the Best Decides strategy.

\begin{table}[!htbp]
    \centering
    \begin{tabular}{|c||c|c||c|c|}
    \hline Model & $s(s_1,s_2)$ & $s(s_1,s_2,s_3)$ &  Mean & Multiplier \\ \hline\hline
     RR & $\frac{s_1+s_2}{2}$ & $\frac{s_1+s_2+s_3}{3}$  & arithmetic  & 1 \\\hline
     Vot & $\frac{s_1+s_2}{2}$ & $\frac{s_1+s_2+s_3}{2}$  & arithmetic  & $\approx 0.8 \sqrt{n}$ \\\hline
     BD & $\max(s_1,s_2)$ & $\max(s_1,s_2,s_3)$  & maximum  & $1$ \\\hline
     WCS & $\frac{s_1+s_2}{\sqrt{2}}$ & $\frac{s_1+s_2+s_3}{\sqrt{3}}$  & arithmetic  & $\sqrt{n}$ \\\hline
     DSS & $\sqrt{s_1^2+s_2^2}$ & $\sqrt{s_1^2+s_2^2+s_3^2}$  & quadratic  & $\sqrt{n}$ \\\hline
     TW & $s_1+s_2$ & $s_1+s_2+s_3$  & arithmetic  & $n$ \\\hline
\end{tabular}
    \caption{Models summary for the six considered models of Sec.\ref{s:communication}. For each model there is given explicit formula for two and three members. In each model the $s_{model}$ has the general form $\mbox{multiplier}\times\mbox{mean}$.}
    \label{tab:cns}
\end{table}

For schemes of aggregation  (Tab. \ref{tab:cnsaggreg}), we obtained two interesting results. First, most of the models we investigated are not affected by the gradual aggregation of information. Second, for models that are affected, the optimal solution is to aggregate information in the smallest possible groups, i.e., in $g=2$ for Weighted Confidence Sharing and $g=3$ for Voting. 

\begin{table}[!htbp]
    \centering
    \begin{tabular}{|c|c||c|c|c|}
    \hline Model & $g$ & Shallow hierarchy &  Random hierarchy & Deep hierarchy \\ \hline\hline
     Vot & 3 & $n^{0.37} $ & $1.22 n^{0.25}$ & $2.00$ \\\hline
     Vot & 4 & $n^{0.16} $ & $1.15 n^{0.08}$ & $1.36$ \\\hline
     Vot & 5 & $n^{0.35} $ & $1.38 n^{0.19}$ & $2.15$ \\\hline
     WCS & 2 & $n^{0.5}$  & $1.13 n^{0.41}$  & $2.41$ \\\hline
     WCS & 3 & $n^{0.5}$  & $1.25 n^{0.37}$  & $2.73$ \\\hline
     WCS & 4 & $n^{0.5}$  & $1.37 n^{0.33}$  & $3.00$ \\\hline
     WCS & 5 & $n^{0.5}$  & $1.48 n^{0.31}$  & $3.23$ \\\hline
\end{tabular}
    \caption{Summary of information-aggregation results (see Sec.~\ref{s:aggregation}) in groups of $g$ agents for affected models, i.e., Voting and Weighted Confidence Sharing. For each model there are provided asymptotic multipliers for three different information-aggregation hierarchies. In each model the $s_{model}$ has the form $\mbox{multiplier}$ times arithmetic mean. Note that for Voting grouping in $g=4$ is very ineffective (as, in fact, it effectively uses the opinions of three out of four participants). Also note that, asymptotically, the most effective approach (i.e., the best for very large groups) for the Shallow and Deep aggregation schemes  is to gather information in the smallest possible groups of agents (i.e., in $g=3$ for Voting and $g=2$ for WCS).}
    \label{tab:cnsaggreg}
\end{table}

It is possible that participants' strategies vary from trial to trial. In such situations, the outcome would be a mixture of strategies (with weights $w_{model}$), that is
\begin{align}
P_{eff}(c) &= \sum_{models} w_{model} P_{model}(c),\\
s_{eff} &= \sum_{models} w_{model} s_{model}.
\end{align}
To distinguish between models, the sole analysis of the group performance might not be enough, as (psychologically) different models of problem-solving can yield  the same performance.  One can test modified schemes that put additional constraints on participants' interactions to investigate communication directly. For example, contact with other members could be limited to voice or text chat, or no feedback may be provided. In addition, participants might be asked to express their confidence explicitly on a Likert scale. However, further experimental work should be carried out to clarify if and when confidence is subjectively accessible and can be communicated explicitly, and when it can be read from participants' behaviors. Preliminary results \citep{Bahrami2012} seem to suggest that the latter is common. Also the amount of feedback could range from full information about the stimulus to simple information about accuracy, to no feedback at all. As a reference, it may serve to examine Social Decision Scheme Theory \citep{Davis1973}, wherein the group decision is considered to be a function of individual choices, regardless of skills, confidences or the difficulty of the task. 

In all the models, interaction is beneficial for the overall performance, except for in the Random Responder model (where the performance is the same as the averaged performance of each individual). It is possible that beyond a certain critical size, groups start to perform worse \citep{Grofman1978}. The models we consider do not predict such a collapse, as they are based on information sharing and do not incorporate phenomena related to motivation and social or technical ability to work in groups.

\section{Conclusion}\label{s:conclusion}

In the paper, we examined mathematical models for solving a two-choice discriminative task by a group of participants. We were interested in how group performance depended upon the performance of the individuals, their ways of communication and their modes of decision aggregation. As a measure of performance, we used the slope of the psychometric function \eqref{eq:slopesigma}, which indicates how performance changes with the difficulty of the task. The higher the slope of $s$ is, the better the performance of the individual (or the group).

We analyzed a number of possible models of decision integration in a joint task. As we moved from 2-person to $n$-person groups, we also had to take into account patterns of interaction among members. Obviously, the choice of the way in which aggregate decisions of group members are made is not always unconstrained. Some of the models, it seems, can be adopted in almost all group decision situations (such as the Random Responder model and the Voting model). Regardless of the properties of the stimuli, people can make their own decisions and vote. For the Best Decides model, we need to assume that the group possess information about the members' performances (e.g., from the feedback). Other models (i.e., the Weighted Confidence Sharing, the Direct Signal Sharing and the Truth Wins models) make direct assumptions about the problem structure or the information that can be shared. Consequently, they can be considered only in particular tasks, in which a certain level of confidence in an individual's own answer can be reached. Our list of models is by no means exhaustive.

We need to be aware of the fact that the presented models are valid only for our specific situation (collaborative decisions in a two-choice perceptive task wherein difficulty can be smoothly adjusted). Other tasks may be analyzed within the same paradigm, such as integrating information in an individual's mind. Several exposures to the same stimulus by a single person, perhaps using different senses or with different noise levels, would be another subject for further investigation. Such an approach is presented in experiments on sensory integration, e.g., by \citet{Ernst2002}, which serve as one of the motivations for the \citet{Bahrami2010} models. Perhaps collaborative decisions in other two-choice tasks (e.g., verbal or mathematical decisions) could also be treated in a similar fashion. However, for many other settings, more advanced models are needed, e.g., ones that take into account more choices or the dynamic interaction between solving a problem in an individual's mind and communicating that decision to the other participants. Nevertheless, we believe that the first step should be to experimentally verify the predicted results of this paper (with an emphasis on the scaling of the performance), before proceeding to more advanced theoretical models. 

\section*{Acknowledgements}

The work was supported by EC EuroUnderstanding grant {\it DRUST} to JRL, Spanish MINCIN project FIS2008-00784 (TOQATA) and ICFO PhD scholarship to PM, and the Polish Ministry of Education and Science (grants: N301 159735, N518 409238) to DP.

\appendix

\section{Approximations}\label{s:approx}

$P(c)$ can be expanded in Taylor series of $c$ around $c=-b$.
\begin{align}
P(c) &= P\left[ -b + (c+b) \right]\\
&= P(-b) + (c+b) P'(-b) + \tfrac{(c+b)^2}{2} P''(-b) + \tfrac{(c+b)^3}{6} P'''(-b) + \ldots
\end{align}
where $P^{(i)}(-b)$ can be found explicitly using \eqref{eq:pdef},
\begin{align}
P^{(i)}(c) = \frac{1}{\sigma^i}H^{(i)}(\tfrac{c+b}{\sigma}). 
\end{align}
In particular $H(0)=1/2$, $H'(0)=1/\sqrt{2\pi}$, $H''(0)=0$, $H'''(0)=-2/\sqrt{2\pi}$.

Consequently,
\begin{align}
P(c) = \frac{1}{2} +  \frac{(c+b)}{\sqrt{2\pi}\sigma} + O\left[ (\tfrac{c+b}{\sigma})^3 \right]\label{eq:pcwitherror},
\end{align}
that is, the approximation error of taking the linear approximation is of the order $(c+b)^3/\sigma^3$ as the quadratic term vanishes.  Plugging $c=0$ we obtain
\begin{align}
P(0) &= \frac{1}{2} + \frac{b}{\sqrt{2\pi}\sigma} + O\left[ (\tfrac{b}{\sigma})^3 \right]\\
& = \frac{1}{2} + s b + O[ (s b)^3 ] 
\end{align}
and similarly, the derivative of \eqref{eq:pcwitherror} in $0$ is
\begin{align}
P'(c)|_{c=0} &= \frac{1}{\sqrt{2\pi}\sigma} + \frac{1}{\sqrt{2\pi}\sigma} O\left[ (\tfrac{b}{\sigma})^2 \right]\\
& = s \left[ 1 + O(s^2 b^2)  \right].
\end{align}
The last equation gives the approximate equation for slope \eqref{eq:slopediff}. Another expression
\begin{align}
\frac{P(0)-1/2}{P'(c)|_{c=0}} &= \frac{b + b O[ (s b)^2 ] }{1+ O[ (s b)^2 ]} = b \left[ 1 + O(s^2 b^2) \right]
\end{align}
yields in the approximate equation for bias \eqref{eq:biasdiff}.

\section{Voting}\label{s:appvot}

\begin{align}
P_{Vot}(c) &= \sum_{k=1}^{\lfloor \frac{n-1}{2} \rfloor} \sum_{\vec{i}} \left[ 1-P_{i_{1}}(c)\right]\cdots \left[ 1-P_{i_{k}}(c)\right] P_{i_{k+1}}(c)\cdots P_{i_{n}}(c)\\
 & +\left[ \frac{1}{2} \sum_{\vec{i}} \left[ 1-P_{i_{1}}(c)\right]\cdots  \left[ 1-P_{i_{n/2}}(c)\right] P_{i_{n/2+1}}(c)\cdots P_{i_{n}}(c)  \right]_{\mbox{if $n$ is even}},\nonumber
\end{align}
After plugging the linearization \eqref{eq:pcwitherror} in the above, and using $\mu_i=s_i(b_i+c)$, each part has the form of
\begin{align}
&\left[ \tfrac{1}{2} - \mu_{i_{1}}+ O( \mu_{i_1}^3 ) \right] \cdots \left[ \tfrac{1}{2} - \mu_{i_k} + O( \mu_{i_k}^3 ) \right]\\
&\times \left[ \tfrac{1}{2} + \mu_{i_{k+1}} + O( \mu_{i_{k+1}}^3 ) \right] \cdots \left[ \tfrac{1}{2} + \mu_{i_n} + O( \mu_{i_n}^3 ) \right]\\
= &\tfrac{1}{2^n} - \tfrac{1}{2^{n-1}} \left( \mu_{i_1} + \ldots + \mu_{i_k}  \right) +  \tfrac{1}{2^{n-1}} \left( \mu_{i_{k+1}} + \ldots + \mu_{i_n}  \right)\\
&+ O(\mu_1^2) + \ldots + O(\mu_n^2)
\end{align}
After applying permutations to the main part (i.e., without the error estimation) we get
\begin{align}
\frac{1}{2^n}{n \choose k} + \frac{1}{2^{n-1}}{n \choose k}\left[ -k + (n-k) \right]\frac{\mu_1+\ldots+\mu_n}{n} \\
= \frac{1}{2^n}{n \choose k}  + \frac{n}{2^{n-1}} \left[ -{n-1 \choose k-1} +{n-1 \choose k} \right] \frac{\mu_1+\ldots+\mu_n}{n},
\end{align}
which is easy to be summed. The first component sums to $1/2$. In the second, binomial coefficients cancel pairwise, except for ${n-1 \choose 0-1}=0$ and  ${n-1 \choose \lfloor (n-1)/2 \rfloor}$ leaving only ${n-1 \choose k}$ for $k=\lfloor (n-1)/2 \rfloor$. Consequently, when $n$ is odd, one gets
\begin{align}
P_{Vot,odd}(c) &= \frac{1}{2} + \frac{n}{2^{n-1}} {n-1 \choose (n-1)/2} \frac{\mu_1+\ldots+\mu_n}{n} + O(\mu_1^2) + \ldots + O(\mu_n^2)
\end{align}
and for even $n$
\begin{align}
P_{Vot,even}(c) &= \frac{1}{2} + \frac{n}{2^{n-1}} {n-1 \choose (n-2)/2} \frac{\mu_1+\ldots+\mu_n}{n} + O(\mu_1^2) + \ldots + O(\mu_n^2).
\end{align}
After the differentiation one obtains the slope \eqref{eq:slopediff} and the bias \eqref{eq:biasdiff}.

\end{document}